\documentclass[twocolumn,showpacs,preprintnumbers,amsmath,amssymb]{revtex4}
\usepackage{graphicx}
\usepackage{dcolumn}
\usepackage{bm}
\begin{document}
\title{
Bragg-Bose glass phase in vortex states of high-$T_{\rm c}$ superconductors \\
with sparse and weak columnar defects
}
\author{Yoshihiko Nonomura}
\altaffiliation[Also at ]{Lyman Lab.\ of Physics, 
Harvard University, Cambridge, MA 02138}
\email{nonomura.yoshihiko@nims.go.jp}
\author{Xiao Hu}
\affiliation{
Computational Materials Science Center, National Institute 
for Materials Science, Tsukuba, Ibaraki 305-0047, Japan}
\date{\today}
\begin{abstract}
Phase diagram of vortex states of high-$T_{\rm c}$ superconductors 
with {\it sparse and weak} columnar defects is obtained by large-scale 
Monte Carlo simulations of the three-dimensional anisotropic, frustrated 
XY model. The Bragg-Bose glass phase characterized by hexagonal Bragg 
spots and the diverging tilt modulus is observed numerically for the first 
time at low density of columnar defects. As the density of defects 
increases, the melting temperature increases owing to ``selected pinning" 
of flux lines. When the density of defects further increases, the transition 
to the Bose glass phase occurs. The interstitial liquid region is observed 
between these two glass phases and the vortex liquid phase. 
\end{abstract}
\maketitle
In vortex states of high-$T_{\rm c}$ superconductors, columnar 
defects introduced by heavy-ion irradiation are strong pinning centers 
of flux lines, and dramatically enhance the critical current. \cite{Civale}
Nelson and Vinokur \cite{Nelson} analyzed these defects and derived 
a phase transition between the Bose glass (BG) and vortex liquid 
(VL) phases. They also showed that this system can be mapped to 
two-dimensional interacting bosons in a random potential and that 
this phase transition corresponds to a localization transition of 
interacting bosons. In the case of dense and strong columnar 
defects, every flux line is trapped by randomly-distributed 
columnar defects at low temperatures, and the BG phase naturally 
appears. Many numerical studies \cite{Wengel,Lidmar,Vestergren} 
have been made in this case. 

In the case of sparse and strong columnar defects, 
part of flux lines are not trapped by defects, and physical 
properties are more complicated. At low temperatures 
untrapped flux lines are also frozen owing to interactions 
with trapped flux lines, and the BG phase is stable. 
At intermediate temperatures the defects still trap some 
flux lines but other interstitial ones are freely moving 
around, and the resistivity becomes finite. This region is 
called as the interstitial liquid (IL). \cite{Radzihovsky} 
At high temperatures all flux lines are depinned and the VL 
phase appears. It is not clarified yet whether there exists a 
phase transition or merely a crossover between the IL and VL. 
A numerical simulation was attempted \cite{Sen} in this case, 
and the existence of the BG phase was confirmed. Two kinds 
of BG phases for dense or sparse columnar defects are divided 
by the Mott insulator phase at the matching field. Recently 
the transition between these two (strong and weak) BG 
phases was observed numerically. \cite{Nandgaonkar} 

\begin{figure}
\includegraphics[width=70mm]{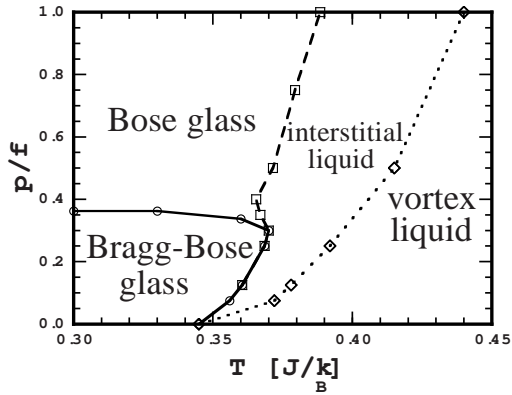}
\caption{\label{fig1}Phase diagram of vortex states 
of high-$T_{\rm c}$ superconductors with sparse 
and weak columnar defects in the plane of number of 
columnar defects per flux line versus temperature. 
$p/f=1$ corresponds to the matching field.
}
\end{figure}

In pure systems, the triangular flux-line lattice (FLL) is formed 
below the melting temperature $T_{\rm m}$. Is such order 
completely destroyed by introducing infinitesimal columnar 
defects? In vortex states with point defects, the long-range order 
of vortex positions is destroyed by infinitesimal defects, but 
the Bragg glass phase \cite{Nattermann,Giamarchi1,Giamarchi2} 
characterized by the {\it quasi long-range order} of vortex 
positions and Bragg spots of the structure factor is stable 
up to a considerable density of point defects. 
In the present Letter, we numerically show that a similar 
phase with Bragg spots associated with hexagonal symmetry 
is also observed in the case of {\it sparse and weak} columnar 
defects. Since the tilt modulus shows a diverging behavior 
in this phase, it corresponds to the Bragg-Bose glass (BBG) 
phase. \cite{Giamarchi2} When the density of columnar defects 
increases, a phase transition to the BG phase takes place. 
We also find that the IL region exists above the BBG or BG 
phases. These results are summarized in Fig.\ \ref{fig1}. 

In this study we perform large-scale Monte Carlo simulations 
based on the three-dimensional anisotropic, frustrated 
XY model on a simple cubic lattice \cite{Li,Hu}: 
\begin{eqnarray}
  \label{XY}
  {\cal H}&=&-\hspace{-0.4cm}
              \sum_{i,j \in ab\ {\rm plane}} \hspace{-0.4cm}
                   J_{ij}\cos \left(\phi_{i}-\phi_{j}-A_{ij}
                                    \right)\nonumber\\
           &&-\frac{J}{\ \Gamma^{2}} \hspace{-0.3cm}
              \sum_{m,n\parallel c\ {\rm axis}}\hspace{-0.3cm}
                   \cos \left(\phi_{m}-\phi_{n}\right)\ ,\\
    A_{ij}&=&\frac{2\pi}{\ \Phi_{0}}\int_{i}^{j}{\bf A}^{(2)}
                                    \cdot {\rm d}{\bf r}^{(2)},
             \ \ {\bf B}={\bf \nabla}\times {\bf A}\ ,
\end{eqnarray}
where the magnetic field is applied along the $c$ axis. The 
periodic boundary condition is applied in all directions. A 
columnar defect is defined as a set of four weak interactions 
$J_{ij}=(1-\epsilon)J$ surrounding a plaquette in the $ab$ 
plane. Such defects distribute randomly in the $ab$ plane 
with probability $p$, and take the same position along the 
$c$ axis. Other interactions in the $ab$ plane are given by 
$J_{ij}=J$. This modeling of columnar defects is a variant 
of our previous study on point defects. \cite{Nonomura1} 
Here we concentrate on the model with the anisotropy 
$\Gamma=5$, the density of flux lines $f=1/25$, the strength 
of defects $\epsilon=0.1$, and the system size $L_{a}=L_{b}=50$ 
and $L_{c}=80$. CPU time is $1\sim 6\times 10^{7}$ Monte Carlo 
steps (MCS) for equilibration, and $1\sim 3\times 10^{7}$ MCS 
for measurements. Since such long simulation time is required 
for taking equilibrium data, the present results are based on 
one typical configuration of columnar defects for each $p$.  

\begin{figure}[t]
\includegraphics [width=7.6cm]{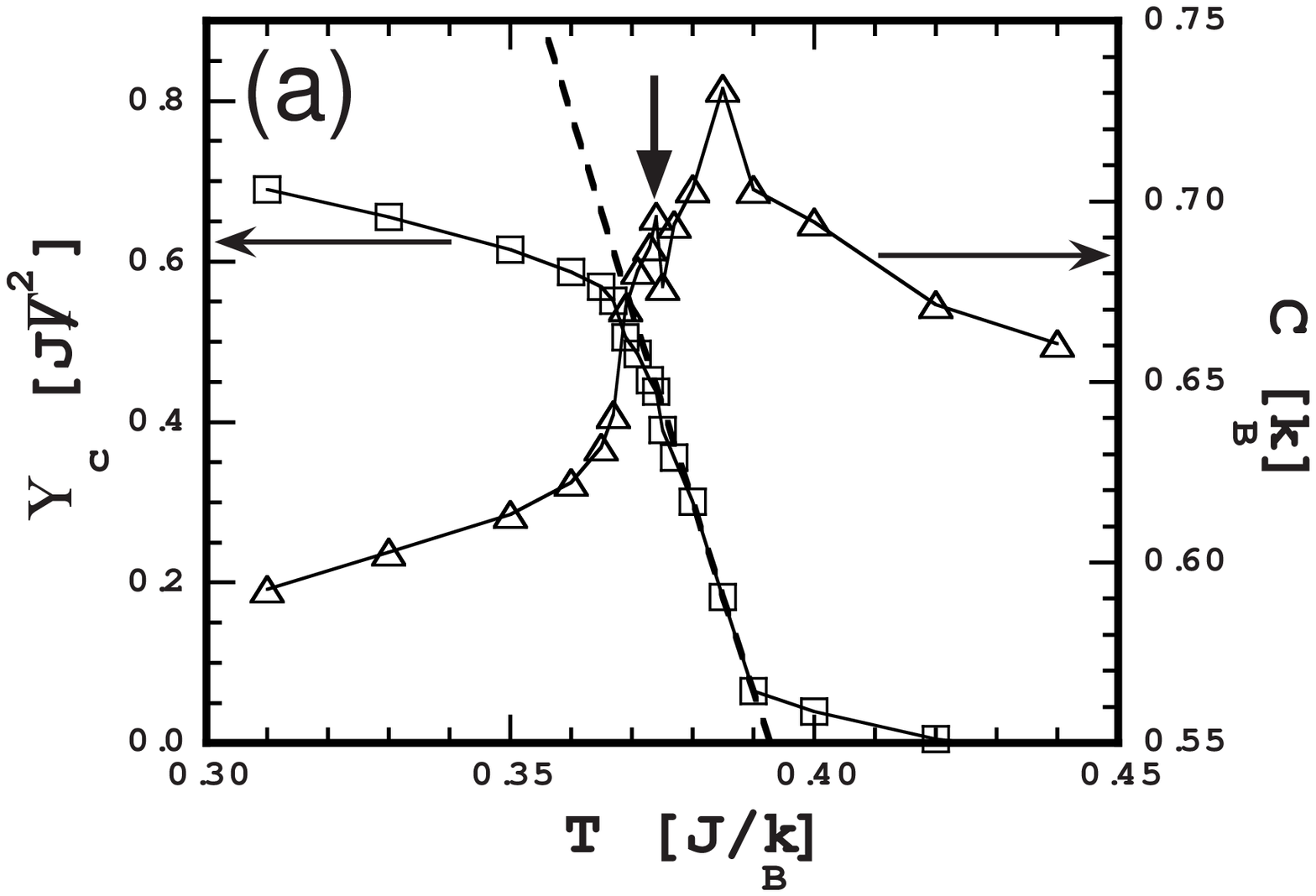}
\hspace*{0.03cm}
\includegraphics[width=7.3cm]{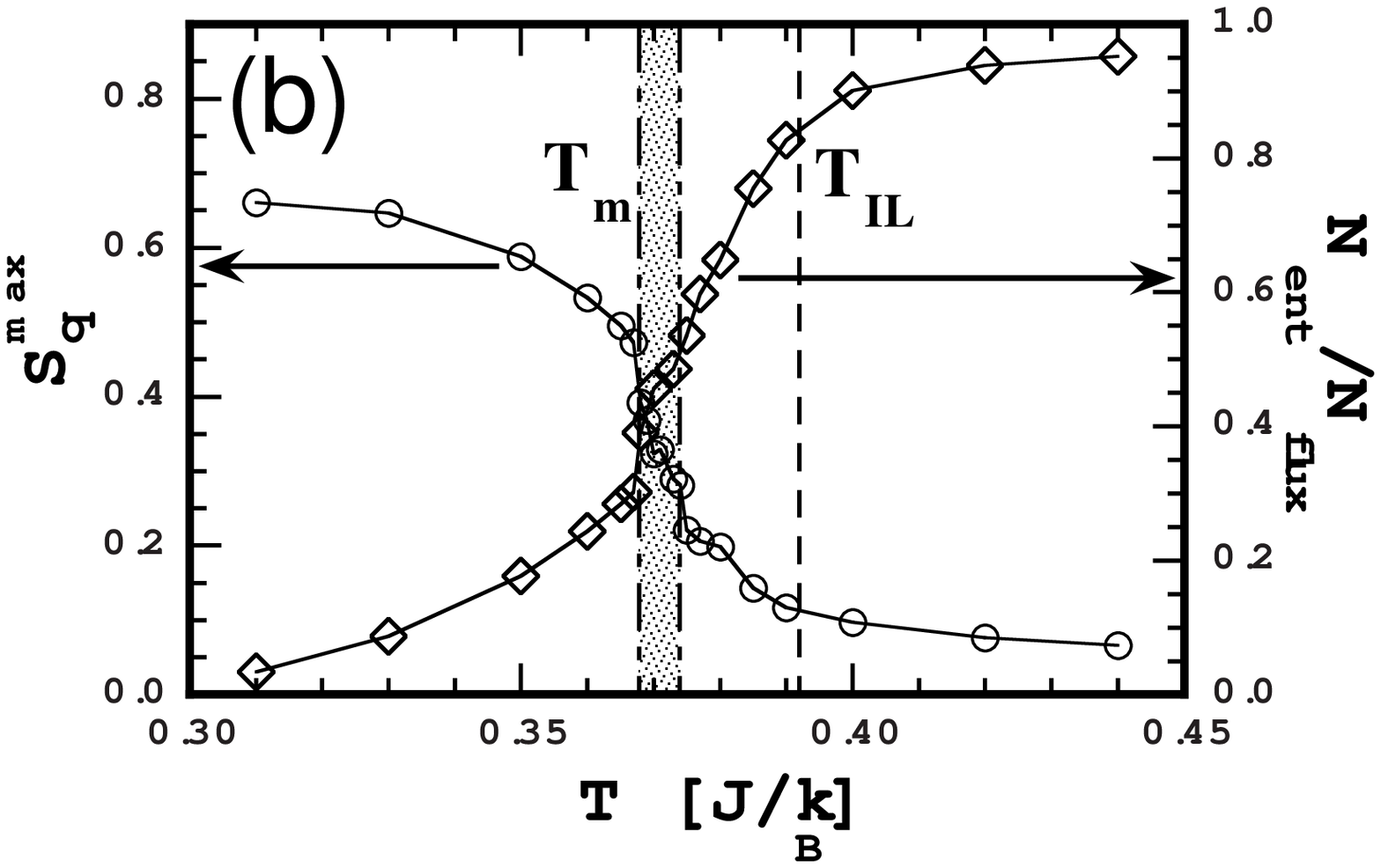}
\includegraphics[width=6.0cm]{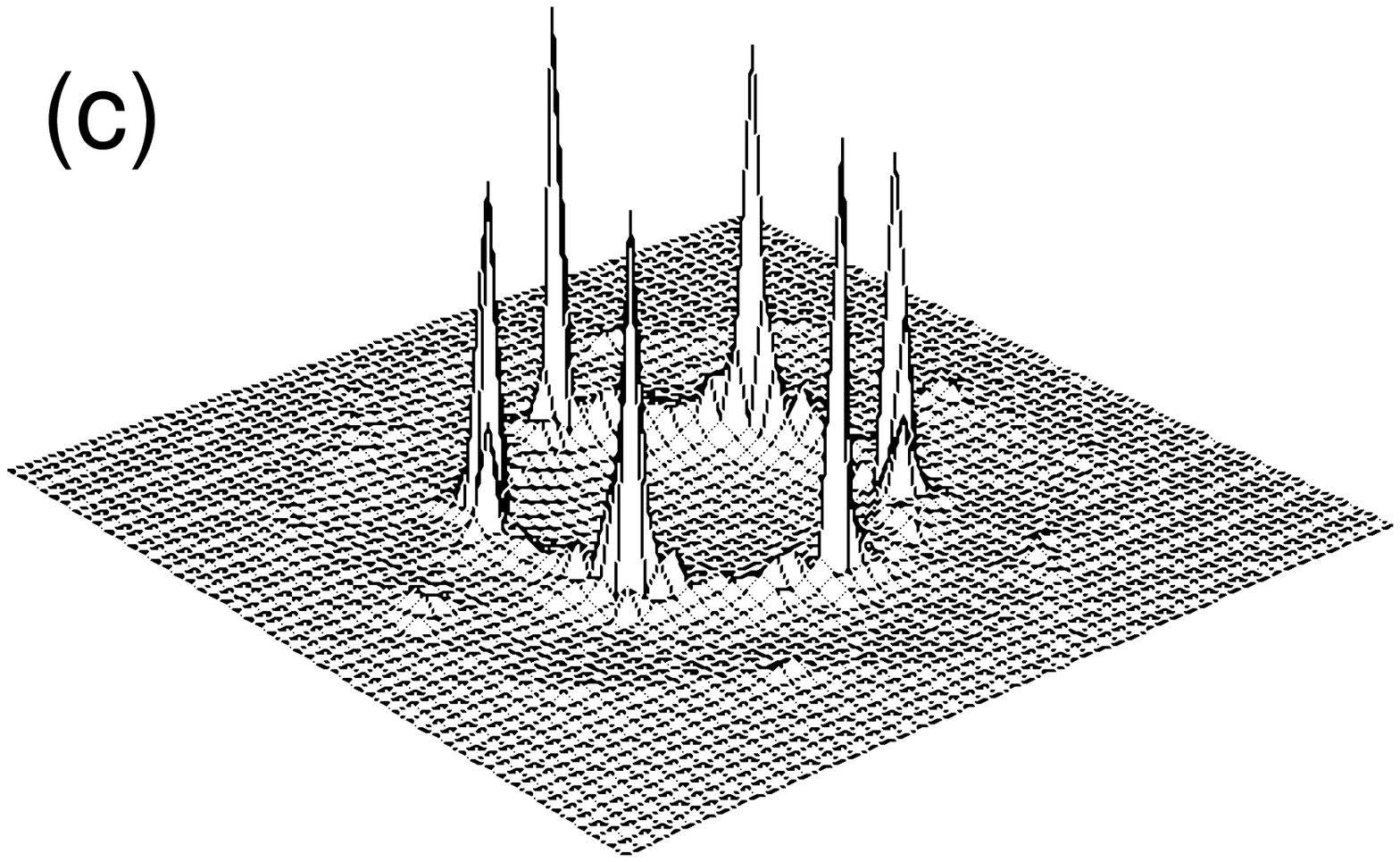}
\caption{\label{fig2}Numerical results for $p/f=0.25$. 
(a) Temperature dependence of the specific heat (triangles) and 
the helicity modulus along the $c$ axis (squares). Marked is the 
sharp peak of the specific heat at $T_{\rm m}$. (b) Temperature 
dependence of the maximum value of the structure factor (circles) 
and the ratio of entangled flux lines to total flux lines (diamonds). 
Coexistence of two phases is observed in the hatched region. 
(c) Structure factor at $T=0.36 J/k_{\rm B}$.
}
\end{figure}
First, we fix the density of columnar defects for $p=0.01$ 
($p/f=0.25$) and vary the temperature. From now on, 
this density is expressed in terms of the number of 
defects per flux line $p/f$ with $f=1/25$ in order to 
clarify the physical situation. The specific heat $C$ 
shows a sharp peak around $T\approx 0.37 J/k_{\rm B}$ 
(marked in Fig.\ \ref{fig2}(a)), and the maximum value 
of the structure factor drops at the same temperature 
(Fig.\ \ref{fig2}(b)). These facts indicate that the melting 
transition occurs at $T_{\rm m}\approx 0.37 J/k_{\rm B}$. 
The structure factor shows sharp hexagonal Bragg spots 
below $T_{\rm m}$ as displayed in Fig.\ \ref{fig2}(c). 
Coexistence of two phases is observed between 
$T=0.368 J/k_{\rm B}$ and $0.374 J/k_{\rm B}$ 
owing to finite-size effects. These findings suggest that 
the transition at $T_{\rm m}$ is of first order. Similar 
results are obtained for $p/f=0.075$, $0.125$ and $0.30$. 

Since the helicity modulus along the $c$ axis 
$\Upsilon_{c}$ takes a finite value when flux lines 
are frozen along the $c$ axis, the vanishing point 
of $\Upsilon_{c}$ corresponds to the boundary 
between the IL and VL. As shown in Fig.\ \ref{fig2}(a), 
temperature dependence of this quantity changes 
around $T_{\rm m}$, and it vanishes at 
$T_{\rm IL}\approx 0.392 J/k_{\rm B}$, which is 
close to the broad peak of $C$. Simulations for 
$p/f=0.075$, $0.125$, $0.50$ and $1.00$ show 
similar temperature dependence of $\Upsilon_{c}$ 
except for monotonic increase of $T_{\rm IL}$ as 
$p$ increases. Size dependence of $T_{\rm m}$ and 
$T_{\rm IL}$ is checked for $p/f=0.075$ and $0.25$ 
using data with $L_{c}=54$, $80$ and $160$, and 
found to be smaller than that of $T_{\rm m}$ of 
the pure system. \cite{Nonomura2} 

As displayed in Fig.\ \ref{fig2}(b), the ratio of entangled 
flux lines to total flux lines $N_{\rm ent}/N_{\rm flux}$ 
rapidly increases from $0.3$ to $0.5$ around $T_{\rm m}$ 
as in the pure system. \cite{Nonomura2} Gradual increase 
of $N_{\rm ent}/N_{\rm flux}$ in the IL region indicates the 
process that untrapped flux lines are freed from trapped ones. 
$N_{\rm ent}/N_{\rm flux}\simeq 0.85$ at $T_{\rm IL}$ is 
comparable to the value at the onset of the VL phase 
in the pure system. \cite{Nonomura2}

\begin{figure}
\includegraphics[width=8.0cm]{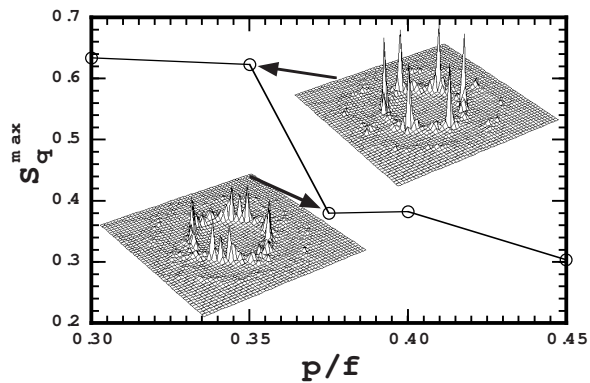}
\caption{\label{fig3}Defect density dependence of the maximum 
value of the structure factor for $T=0.300 J/k_{\rm B}$. Structure 
factors at $p/f=0.35$ and $0.375$ are also displayed. 
}
\end{figure}
Second, we vary the density of columnar defects for 
$T=0.300 J/k_{\rm B}$. As shown in Fig.\ \ref{fig3}, 
the maximum value of the structure factor drops at 
the BBG-BG phase boundary. Although anomalies of 
thermodynamic quantities such as the internal energy 
are too weak to observe numerically in the present 
study, the change of the symmetry of Bragg spots is 
quite apparent as depicted explicitly in Fig.\ \ref{fig3}. 
The hexagonal symmetry of Bragg spots at $p/f=0.35$ 
(BBG phase) is lost at $p/f=0.375$ (BG phase) and a 
glassy ring-like pattern appears. Similar results are 
obtained for $T=0.330 J/k_{\rm B}$ and $0.360 J/k_{\rm B}$. 

\begin{figure}
\includegraphics[width=7.0cm]{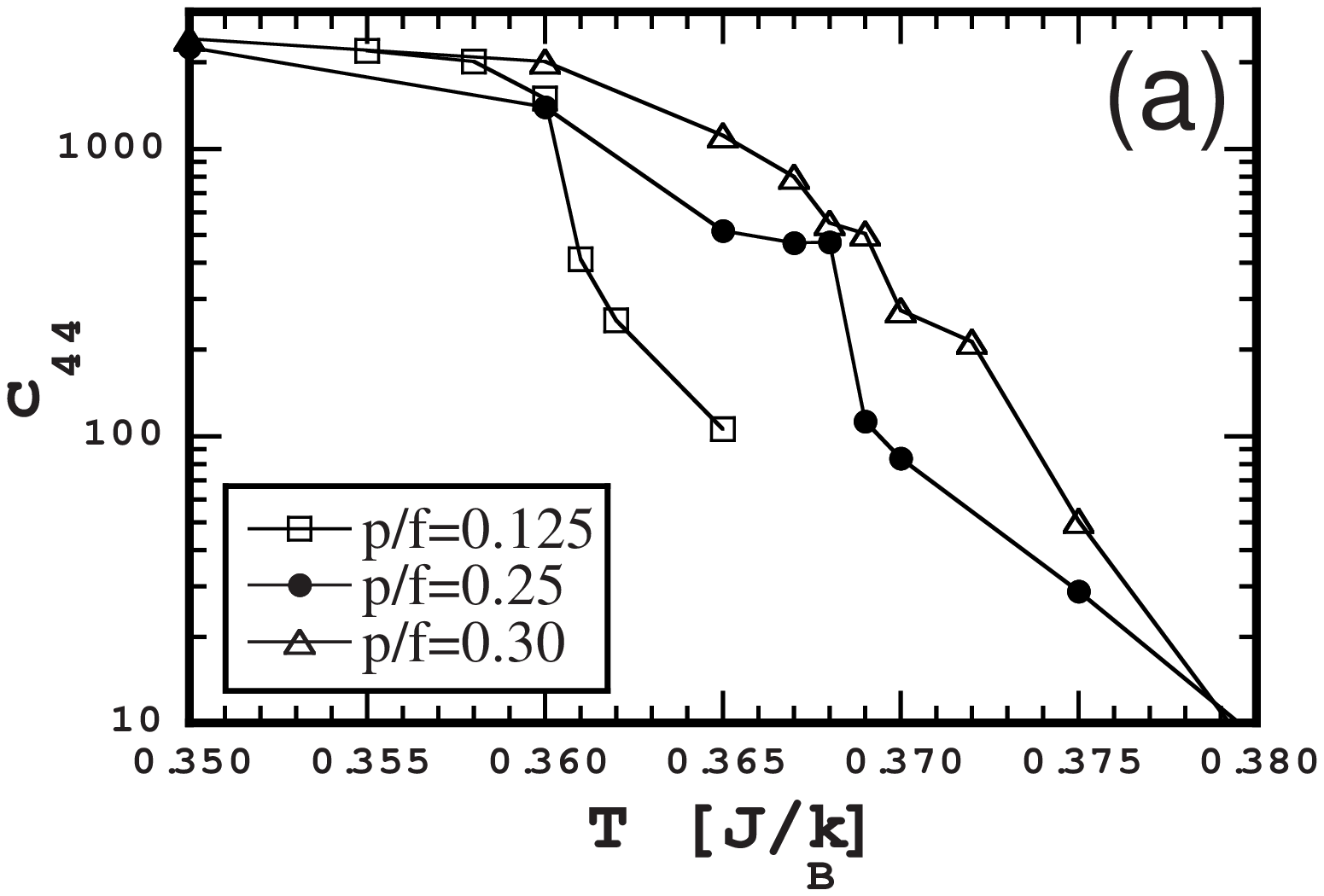}
\includegraphics[width=7.0cm]{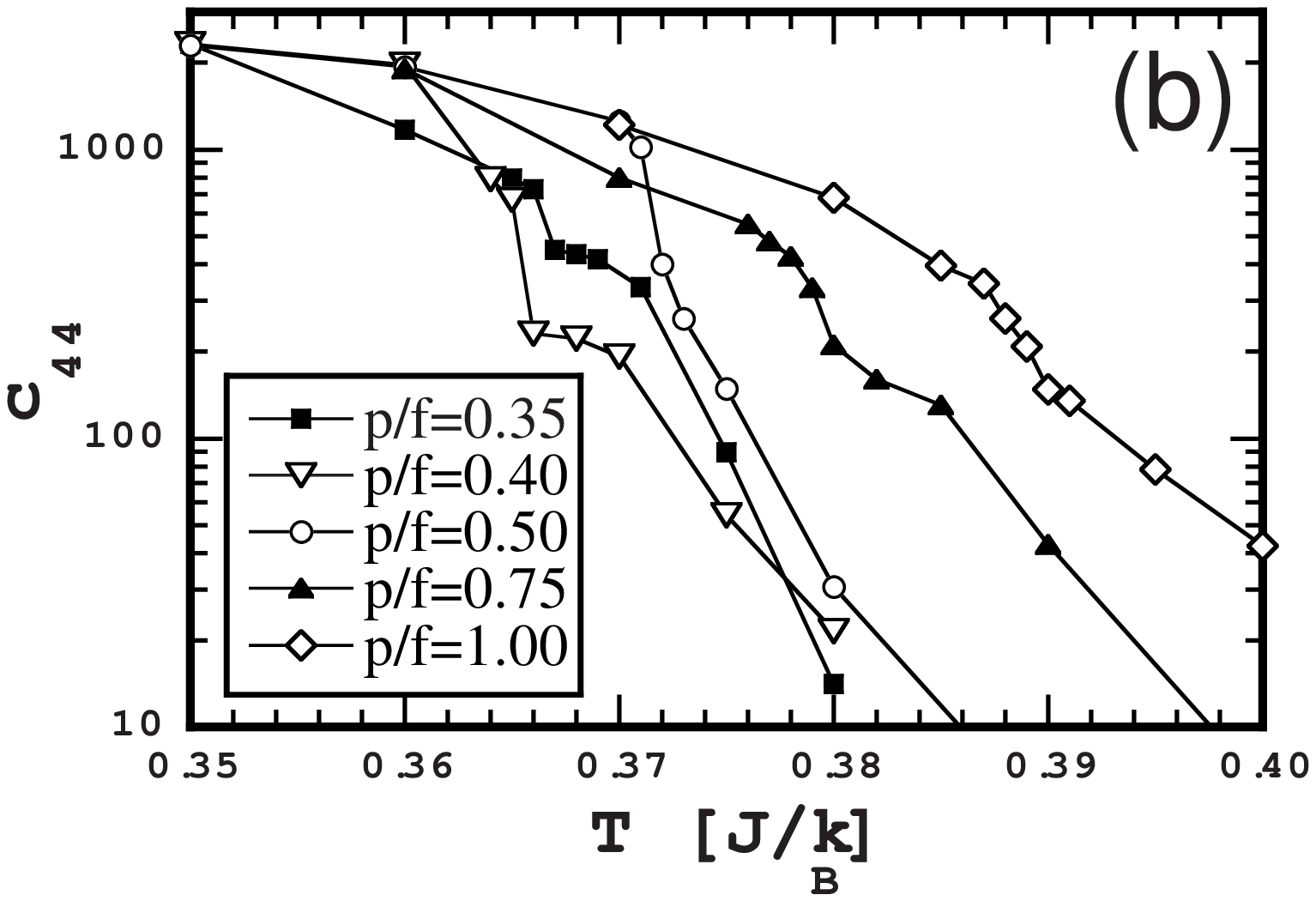}
\caption{\label{fig4}Temperature dependence of the tilt 
modulus defined in eq.\ (\ref{tilt}) for (a) $p/f=0.125$ 
to $0.30$ and for (b) $p/f=0.35$ to $1.00$.
}
\end{figure}
Third, in order to evaluate the BG transition temperature 
$T_{\rm g}$, the tilt modulus $c_{44}$ is measured. 
The inverse of this quantity is proportional to the 
zero-field susceptibility of the transverse magnetization 
$m^{\perp}$ for the transverse field $H^{\perp}$, namely 
$c_{44}^{-1}\sim \left.\partial\langle m^{\perp}\rangle
                   /\partial H^{\perp}\right|_{H^{\perp}=0}$. 
The Hamiltonian under the transverse field is given by 
${\cal H}={\cal H}_{0}-H^{\perp}m^{\perp}$,
where ${\cal H}_{0}$ stands for the Hamiltonian without 
the transverse field given in eq.\ (\ref{XY}), and the 
transverse magnetization $m^{\perp}$ is defined from 
the sum of the difference of positions of flux lines 
along the direction of $H^{\perp}$ at the top and 
bottom $ab$ planes. Explicitly, $c_{44}$ is given, 
aside from the coefficient, by \cite{Lidmar}
\begin{equation}
  \label{tilt}
  c_{44}\sim\frac{N_{\rm flux}k_{\rm B}T}
                 {\left.\langle\left(m^{\perp}\right)^{2}
                        \rangle\right|_{H^{\perp}=0}}\ .
\end{equation}
Temperature dependence of $c_{44}$ for $p/f=0.125$ to 
$0.30$ (BBG-IL boundary) and for $p/f=0.35$ to $1.00$ 
(BG-IL boundary) is displayed in Figs.\ \ref{fig4}(a) and 
\ref{fig4}(b), respectively.  In both cases, $c_{44}$ shows 
a large jump (note that these figures are drawn in semi-log 
scale) at $T_{\rm g}$. This jump is expected to be infinite 
in the thermodynamic limit. Since $T_{\rm g}$ coincides 
with $T_{\rm m}$ for $p/f \leq 0.30$ (see Figs.\ \ref{fig1} 
and \ref{fig4}(a)), the diverging behavior of $c_{44}$ is 
observed in the whole region of the BBG phase. 

\begin{figure}
\includegraphics[width=8.6cm]{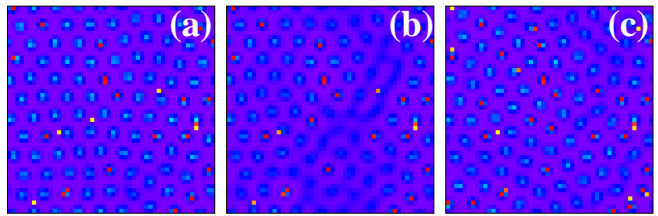}
\caption{\label{fig5}Top view of averaged positions of flux lines 
with respect to $80$ planes and $2000$ independent snapshots 
at (a) $p/f=0.25$ and $T=0.360 J/k_{\rm B}$ (BBG phase), 
(b) $p/f=0.25$ and $T=0.380 J/k_{\rm B}$ (IL region), and 
(c) $p/f=0.40$ and $T=0.360 J/k_{\rm B}$ (BG phase). 
Columnar defects are painted in yellow (no flux lines) to red 
(filled with flux lines), and other regions are painted in violet 
(no flux lines) to light blue (largest probability of flux lines). 
}
\end{figure}
Next, real-space images of flux lines are shown in 
Fig.\ \ref{fig5}. Positions of flux lines are averaged 
in all $ab$ planes and during $10^{6}$ MCS (measured 
for every $500$ MCS). Details of these rainbow-color 
figures are explained in the caption. At $p/f=0.25$ 
and $T=0.360 J/k_{\rm B}$ (Fig.\ \ref{fig5}(a)), all flux 
lines are localized and seem to form a triangular lattice, 
which characterizes the BBG phase. At $p/f=0.25$ and 
$T=0.380 J/k_{\rm B}$ (Fig.\ \ref{fig5}(b)), the overall 
order does not exist any more, though columnar defects still 
trap flux lines. These properties are consistent with those 
of the IL region. At $p/f=0.40$ and $T=0.360 J/k_{\rm B}$ 
(Fig.\ \ref{fig5}(c)), although all flux lines are localized, the 
FLL structure seen in Fig.\ \ref{fig5}(a) is totally destroyed. 
Since most columnar defects are occupied by flux lines 
unless other occupied ones are nearby, many dislocations 
inevitably appear. These properties signal the BG phase. 

Finally, interesting points of the phase diagram given in 
Fig.\ \ref{fig1} are discussed. The most important feature 
is the existence of the BBG phase, which has not been 
observed experimentally or numerically until present. 
Giamarchi and Le Doussal \cite{Giamarchi3} argued that 
nature of the glass order of the three-dimensional vortex 
states with columnar defects is equivalent to that of the 
two-dimensional vortex states with point defects, and 
pointed out \cite{Giamarchi2} the possibility of the 
BBG phase. Stability of the quasi long-range order 
against topological dislocations was discussed by 
D.~S.~Fisher, \cite{Fisher} and he showed that this 
order is destroyed in $d=2$ but stable in $d=3$. 
Our finding is consistent with these analytic studies. 

The melting temperature $T_{\rm m}$ of the BBG phase 
increases as the density of columnar defects $p$ increases 
up to $p/f=0.30$. This behavior may look strange at a glance 
because the FLL order is enhanced by introducing randomness. 
The key to understand this exotic behavior is the real-space 
image of flux lines in the BBG phase given in Fig.\ \ref{fig5}(a). 
There exists some unoccupied columnar defects (yellow dots) 
even though no occupied defects are nearby. This fact suggests 
that the elastic repulsion between flux lines is dominant in the 
BBG phase and makes the columnar defects far from stable 
positions of FLL remain unoccupied. Owing to this ``selected 
pinning'' mechanism, increase of $p$ does not seriously affect 
energy cost for FLL distortion, but bring energy gain from 
pinning of flux lines by the columnar defects close to stable 
positions of FLL. In this case, $T_{\rm m}$ increases as $p$ 
increases. As temperature increases, the BBG melts into the 
IL and the number of occupied flux lines increases as shown 
in Fig.\ \ref{fig5}(b). This behavior can also be explained 
naturally by the above mechanism. 

When $p$ further increases, energy gain from pinning of flux 
lines by many columnar defects exceeds that from FLL formation, 
and the transition to the BG phase takes place. The position of 
the BBG-BG phase boundary thus depends on the strength 
of columnar defects $\epsilon$. When $\epsilon$ increases, 
this phase boundary shifts toward smaller $p$. Actually, 
simulations for $\epsilon=0.5$ at $p/f=0.25$ results in 
the BG phase at low temperatures, where the BBG phase 
exists for $\epsilon=0.1$ (Fig.\ \ref{fig1}). There might 
be a critical value $\epsilon_{\rm c}$ above which no 
BBG phase exists for any $p$. 

The boundary between the BBG or BG and IL phases (on 
which $c_{44}$ jumps) is not monotonic in the vicinity 
of the BBG-BG phase boundary. This result shows a sharp 
contrast to the cage-model analysis, \cite{Goldschmidt} 
where the ``melting temperature'' monotonically increases 
as the strength of columnar defects increases. In the cage 
model, the system is reduced to a single flux line in a 
mean-field potential, and the BBG and BG phases cannot 
be distinguished any more. We consider that the monotonic 
melting line obtained from the cage model corresponds to the 
BG-IL phase boundary above $p/f=0.40$. When this smooth 
curve is extrapolated to $p=0$, it seems to meet the melting 
temperature of the pure system, $T_{\rm m}^{\rm pure}$. 
The excess superconducting region below $p/f=0.40$ is 
due to a nontrivial interplay between columnar defects 
and FLL structure, which is the origin of the BBG phase. 

In the melting transition of FLL in pure systems, sudden 
entanglement of flux lines and increase of dislocations in the 
$ab$ plane occur simultaneously, and dominance of these 
two contributions has been discussed. Random columnar 
defects along the $c$ axis suppress entanglement of 
flux lines while enhance dislocations in the $ab$ plane. 
The relation $T_{\rm m}>T_{\rm m}^{\rm pure}$ might 
suggest that proliferation of entanglement of flux lines 
rather than that of dislocations in the $ab$ plane is 
the main driving force of the melting transition. 

Three-dimensional vortex systems with columnar defects 
can be mapped to the two-dimensional interacting bosons 
in a random potential in the ground state. \cite{Nelson} 
Then, the present result indicates that the quasi long-range 
order of interacting bosons in two dimensions is stable 
in the presence of sparse and weak randomness. 

In conclusion, the phase diagram of vortex states of 
high-$T_{\rm c}$ superconductors with sparse and weak 
columnar defects is obtained by large-scale Monte Carlo 
simulations as shown in Fig.\ \ref{fig1}. The Bragg-Bose 
glass phase characterized by Bragg spots with hexagonal 
symmetry and the diverging tilt modulus is observed for 
the first time. The melting temperature to the interstitial 
liquid region increases as the density of columnar defects 
increases owing to a ``selected pinning" mechanism. 
When the density of defects further increases, a transition 
to the Bose glass phase takes place. The interstitial liquid 
region lies between the Bragg-Bose glass or Bose glass 
phases and the vortex liquid phase. 

The authors thank D.~R.~Nelson, L.~Radzihovsky, B.~Halperin, 
E.~Zeldov, U.~C.~T\"auber, L.~Balents, R.~Ikeda and A.~Tanaka 
for helpful comments. Numerical calculations were performed 
on Numerical Materials Simulator (NEC SX-5) at Computational 
Materials Science Center, National Institute for Materials Science, 
Japan. This study is partially supported by the Priority Grant 
No.\ 14038240 from Ministry of Education, Culture, Sports, Science 
and Technology, Japan. One of the authors (Y.~N.) was supported 
by the Atomic Energy Research Fund from the same Ministry. 

\end{document}